\documentclass[preprint]{vgtc}               

\ifpdf
  \pdfoutput=1\relax                   
  \usepackage{graphicx}                
  \DeclareGraphicsExtensions{.pdf,.png,.jpg,.jpeg} 
\else
  \usepackage{graphicx}                
  \DeclareGraphicsExtensions{.eps}     
\fi%

\graphicspath{{figures/}{pictures/}{images/}{./}} 

\usepackage{microtype}                 
\PassOptionsToPackage{warn}{textcomp}  
\usepackage{textcomp}                  
\usepackage{mathptmx}                  
\usepackage{times}                     
\usepackage{cite}                      
\usepackage{tabu}                      
\usepackage{booktabs}                  

\onlineid{113}

\vgtccategory{Research}

\vgtcinsertpkg

\preprinttext{\href{http://ieeevis.org/year/2017/info/overview-amp-topics/vis-in-practice}{The IEEE Workshop on Visualization in Practice, 2017} }

\title{Deck.gl: Large-scale Web-based Visual Analytics Made Easy}

\author{Yang Wang\thanks{e-mail: gnavvy@uber.com}\\ %
        \scriptsize Uber Technologies, Inc %
}

\abstract{
In this paper, we demonstrate how deck.gl, an open source project born out of data-heavy visual analytics applications, has grown into the robust visualization framework it is today. We begin by explaining why we built another data visualization framework in the first place. Then, we summarize our design goals (distilled from our interactions with users) and discuss how they guided the development of the framework's main features.  We use two real-world applications of deck.gl to showcase how it can be applied to simplify the creation of data-heavy visualizations. We also discuss our lessons learned as we continue to improve the framework for the larger visualization community.
} 

\CCScatlist{
  \CCScat{K.6.1}{Human-centered computing}{Visualization}{Visualization systems and tools};
  \CCScat{K.7.m}{Human-centered computing}{Visualization}{Visualization theory, concepts and paradigms}
}

\nocopyrightspace

\begin{document}


\firstsection{Introduction}

\maketitle


The advancement of almost every modern domain depends on data. Companies and organizations invest heavily in infrastructure for data storage and processing, but unless they can extract meaning from it, the investment is a sunk cost with little reward.

Visualization, as an effective means of bridging human knowledge and data to drive decisions, has gained popularity in the industry in recent years. Nonetheless, despite the amount of effort being put forth by the community, it is still non-trivial for domain experts and practitioners to create large-scale visual analytics solutions that are also reusable in the long term.

While there are more than a handful of frameworks or toolkits in the wild that one can use to create visualizations, they are either native desktop applications lacking portability and flexibility of collaborating or integrating with other applications; or toolkits that are web-based but not performant enough to handle large-scale data sets; or frameworks that are sufficiently performant but originally designed for domains such as video games or media processing, with overheads and steep learning curve for those unfamiliar with these use cases.

In an attempt to narrow the gap, we present the design and rationales behind \textit{deck.gl}, a web-based visualization framework that is highly performant, yet features an easy-to-grasp visual compositing paradigm. At the core of deck.gl lies a collection of proven layers grown out of several in-house data-heavy applications, each with hundreds of daily users. The layers enable developers and designers to quickly prototype through composition, while the framework offers comprehensive data handling and interaction mechanisms for building production-ready analytical solutions.

Besides these enterprise use case, we also envision deck.gl to serve a purpose as a tool for researchers. As a matter of fact, not all research projects are actively maintained as the authors graduate or switch their focuses. This leads to unnecessary, often duplicated efforts for further development on top of of existing solutions. With scalability and usability in its DNA and the ever-growing catalog of layers selected from real-world use cases, we believe deck.gl is beneficial as a building block for both researchers and practitioners to jump-start their projects. With the ongoing collaborative efforts after open sourcing the project, we envision deck.gl will encourage more practitioners to share and contribute to the visualization community.

In this paper, we first outline the design goals distilled from our daily interactions with internal and external users. We then describe the primitive-instance-layer (PIL) paradigm, which has proven to be easy-to-grasp as we onboard more users. Next, instead of providing a full API manual, we highlight a few select features that make deck.gl stand out. Finally, we demonstrate how to use these features in practice with two real world applications, followed by lessons learned, caveats and trade-offs of building large-scale visualizations.

\section{Design Goals}

Having undergone multiple iterations, deck.gl has grown from a handful of independent layers at its debut into a popular open source framework featuring over 3,500 stars on GitHub. In an effort to keep development on track, we constantly review best practices and conduct longitudinal studies with active GitHub users with related project backgrounds (e.g. GIS, computer graphics, video games, etc.). In this section, we present a summary of the highly regarded features that a visualization framework should embody, explored from three perspectives: Scalability (S), Usability (U), and Extensibility (E).

\textbf{S1: A framework should facilitate visualizing data at scale.} Surprisingly, we found users value scalability over usability. \textit{"I don't mind learning new APIs as long as it unlocks new possibilities."} one user commented. Also, although it is agreed the final visualization design should be concise with task-irrelevant data being filtered out, it is desirable to surface such "noises" in early iterations for pattern recognition.

\textbf{S2: A framework should handle dynamic data, smartly.} Data visualization often involves data transformation, which can be expensive for large, streaming data sets. To reduce redundant computation, a framework should incorporate smart batching and differencing logics to prevent unnecessary data processing, while providing the flexibility for fine grained controls over the procedure.

\textbf{U1: A framework should feature an easy-to-grasp information composition model.} Often, users find it hard to model the mapping between data and visual representations, especially when either the cardinality or the dimensionality of the input domain is large. An easy-to-grasp mental model would help shorten such learning curve for decomposing an existing visualization design and reconstructing a new design that fits their own needs.

\textbf{U2: A framework should come with ready-to-use examples.} When presented with a data set, one common practice for visualization is to seek and apply existing designs for quick prototyping. Thus besides the core functionality, a visualization framework should incorporate functional, standalone examples that allow users to plug in their data and search for salient patterns. Very likely, these examples will not be the final solution out-of-box, but they serve to jump-start the exploration process with little hassle.

\textbf{E1: A framework should be interoperable with other existing solutions.} A visualization framework should be a handy tool to facilitate the problem-solving process. Instead of building an all-in-one solution with a steep learning curve, it is more desirable for users if they can leverage their prior knowledge of existing tools they are more familiar with. The framework should instead focus on addressing any pain points. 

\textbf{E2: A framework should be easily extensible to attract community contributions.} Building a framework is time-consuming and often leads to questions regarding early stage ROI. It is also difficult for a single developer or even a team to dedicate all their efforts to maintaining and pushing the boundaries. Instead, one viable approach is to focus on the extensibility of the framework and try to engage the whole community in contributing and sustaining the effort.

\section{Key Features}
Given the above design goals, we describe the design concepts, the rationale, and a selection of key features of deck.gl. Again, all features are distilled from real-world use cases in our day-to-day work on visualization-aided data-heavy applications.

\subsection{The Primitive - Instancing - Layering Paradigm}
\label{sec:PIL}

In an effort to survey and summarize common practices of architecting visualizations, we introduce the \textit{primitive-instancing-layering} (PIL) paradigm. In a nutshell, when visualizing data we are plotting a set of similar but not necessarily identical visual elements and stacking them as a deck of \textit{layer}s. Each layer has at least one \textit{primitive} model, be it simple geometries such as circles or rectangles, or complicated meshes of arbitrator shapes attached with textures. Given an array of input data, the primitives are \textit{instance}d by mapping the attributes of each datum to visual channels such as position, size, color, angle, etc.

\begin{figure}[tb]
  \centering
  \begin{minipage}{0.238\textwidth}
    \centering
    \includegraphics[width=\textwidth]{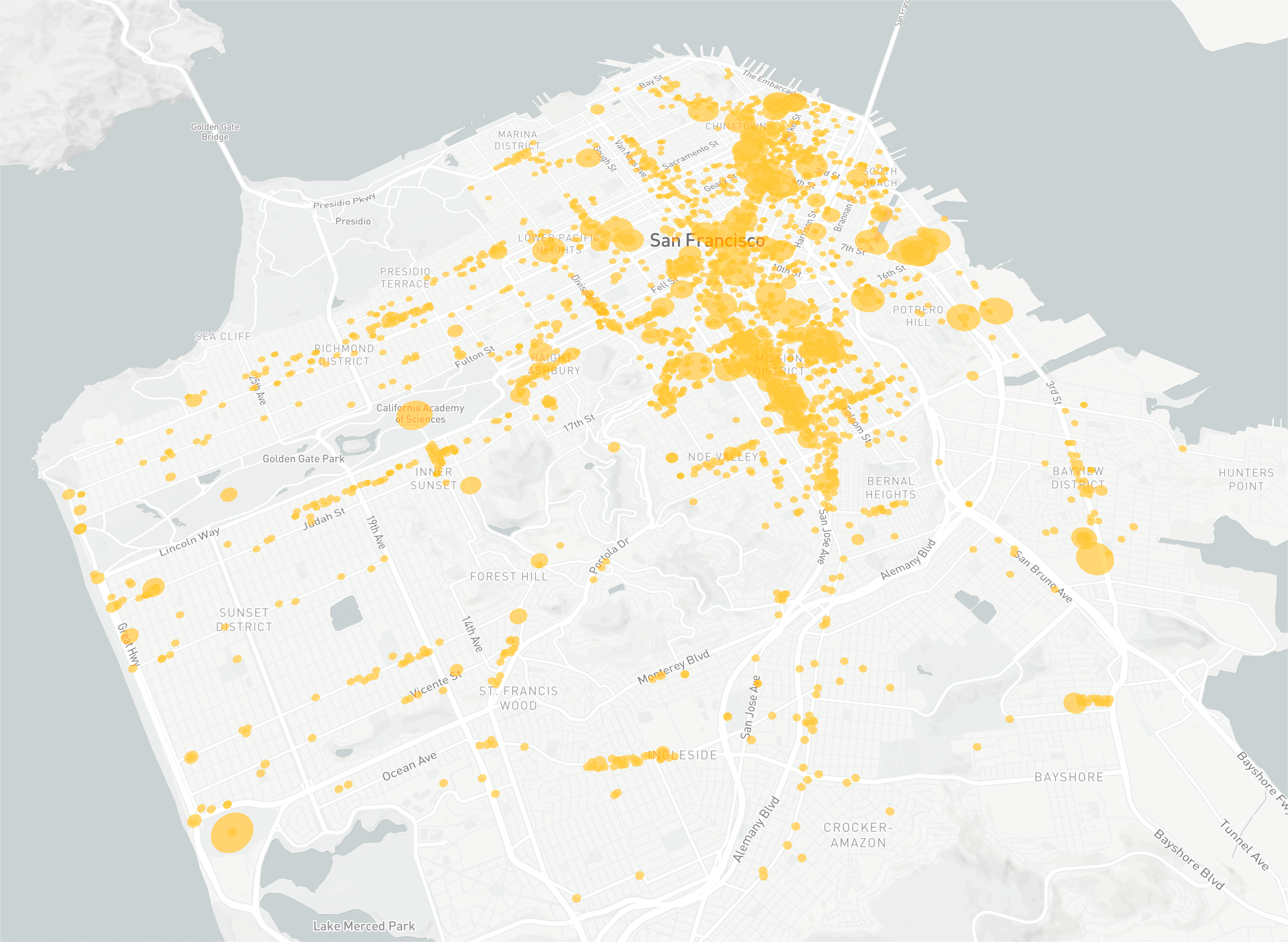}
    \vspace{-5mm}
  \end{minipage}\hfill
  \begin{minipage}{0.238\textwidth}
    \centering
    \includegraphics[width=\textwidth]{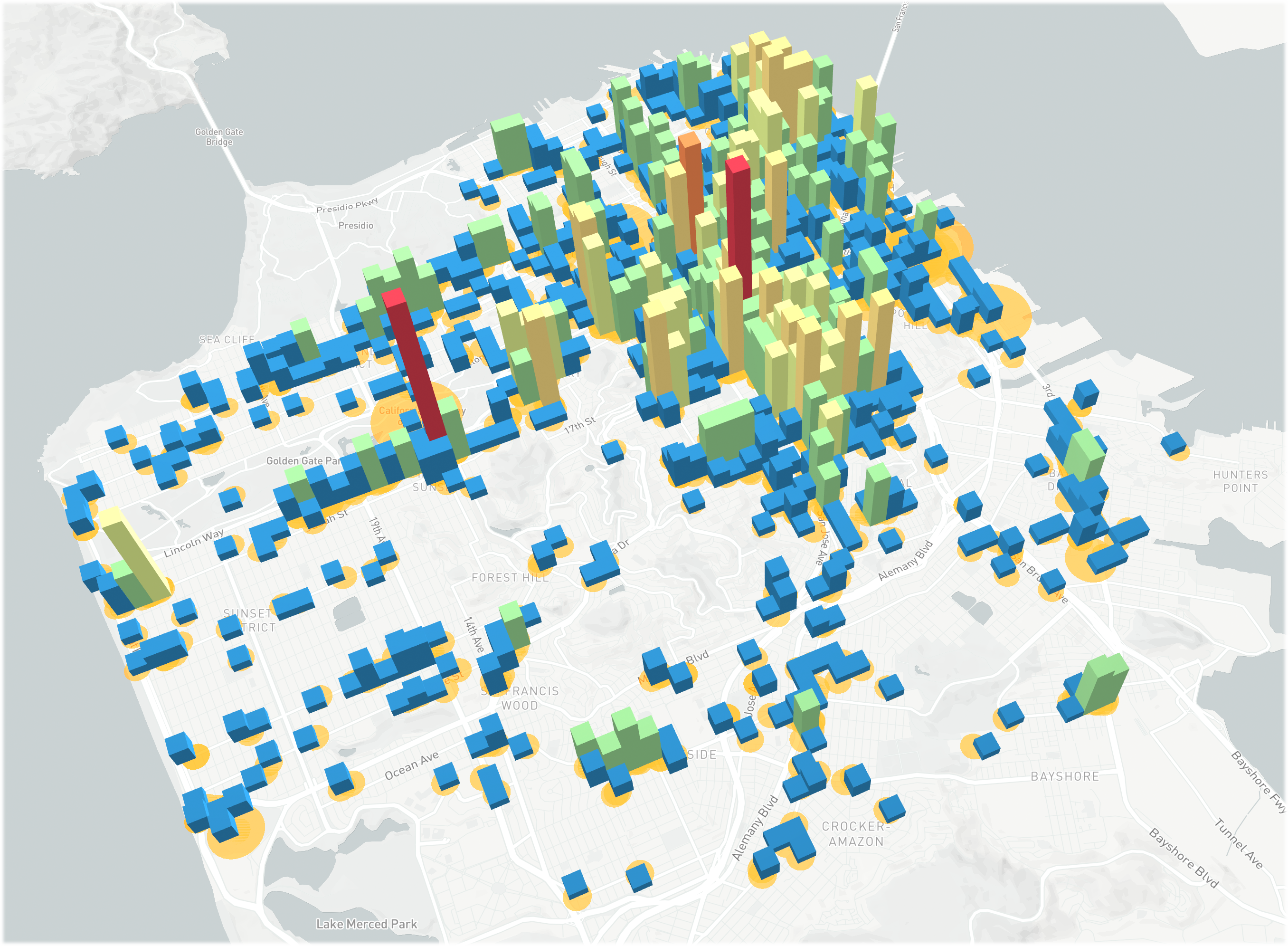}
    \vspace{-5mm}
  \end{minipage}
  \caption{An example visualization showing the distribution of bike parking spots in downtown San Francisco. Left: scatterplot layer with circle as the primitive; Right: extruded grid layer with cuboid as the primitive.}
  \label{fig:PIL}
\end{figure}

For example, in the San Francisco bike parking spots visualizations depicted in Figure \ref{fig:PIL}, the bottom of each view is a map layer providing the geospatial context. On the left, the bike parking spots are visualized using a basic scatterplot layer with circles as the primitive geometry. The circles are instanced by projecting the input latitude and longitude coordinates of each parking spot to screen space positions and sized by the usage frequencies of each parking spot. Similarly, we overlay another extruded grid layer on top of the scatterplot layer to show usage by area in the view to the right. The primitive of the extruded grid layer is a cuboid. Each cuboid instance is placed by their positions, and color and height correspond to the aggregated number of parking spots within the grid cells.

As we can see, users can easily create visualization by stacking layers with desired primitives (\textbf{U1}). In addition to intuitiveness, the PIL paradigm also fits perfectly with the instanced rendering functionality in WebGL, which is now ubiquitously supported by modern browsers. Instanced rendering executes the same drawing commands many times in a row, with each producing a slightly different visual element. This can be very efficient when rendering a large number of glyphs with very few API calls, which is key to interactive visualization at scale compared to regular in-browser visualizations that generate a DOM tree.

\subsection{Built-in Layers}

The whole deck.gl framework is built on top of the layer composition concept. To satisfy ever growing application requests, we have grown the built-in layers from a handful of core primitive layers to composite layers that either contain more than one primitive model or render to multiple sublayers, as well as offscreen functional layers for post-processing of data filtering, aggregation, advanced layout, and interaction support.

\subsubsection{Composite Layers}

Besides the primitive layers that render one geometry primitive at a time, one routine use case we found as we onboard more users and applications is that users often try to render a collection of primitive layers together. One example of this is the GeoJSON layer. The GeoJSON layer takes in a GeoJSON \cite{butler2008geojson} formatted file, but instead of rendering as one primitive layer, it checks the geometry types from the input file, extracts features sharing the same primitive type, and delegates the rendering to the corresponding primitive layers. As such, users only need to pass in the GeoJSON files, and deck.gl will take care of the rendering for them in a performant way.

\subsubsection{Functional Layers}

Another type of layer are the functional layers. Unlike primitive layers, functional layers are often invisible and are used as an intermediate buffer together with the input data to provide context-aware rendering. We give an example in Section \ref{sec:gpgpu} to show how to use a instanced cylinder layer, together with a scatterplot layer to achieve nearest point picking with little extra effort.

In retrospect, community feedback as well as our own experience suggest that the layering concept not only made architecting information easier, but also led to componentization of reusable layer modules, which helped bring in community contributions. (\textbf{E2})

\subsection{Beyond Declarative API}
\label{sec:declaritive-api}

Like most modern web frameworks (e.g. \cite{react.js} \cite{aframe}), deck.gl wraps imperative an API with a declarative one.

The declarative API leads to a better usability of the framework. Compared to asking for explicit instructions describing \textbf{how} to manipulate data, update buffers, trigger rerenderings, etc., the declarative API of deck.gl only asks for \textbf{what} to present, and frees users from worrying about low-level mutative details. As a result, users no longer have to maintain a mental model of the various visual states mutated by data or interactions. They can simply expect that any property changes will cause a complete re-generation of a new visualization representing the desired visual state. In essence, this results in a stateless mapping and what users declare will lead to what they see. As such, the mapping between the code and the resulting visualization is much easier to reason about.

The increase in usability, however, posits challenges of internal imperative API implementations. Recall we are designing a visualization framework capable of handling data at scale, and the key to making data-heavy applications performant lies in minimizing redundant computations. To achieve this, we designed the data wrangling logics for deck.gl layers at three granularity levels. (\textbf{S2})

\textbf{Per-layer update:} If a new data object is passed to the layer that causes the shallow equality check to fail, we invalidate all attributes, discard their vertex buffers from previous calculations and re-generate them from scratch. If a data object has changed but the reference remains the same, we ignore the changes, and reuse the existing vertex buffers to prevent deep comparisons, which is expensive.

\textbf{Per-attribute update:} Re-generating vertex buffers for all layer attributes can be expensive, especially for layers of large memory footprints. Often, users only want to conditionally update one or a few attributes. For example, changing the color of the given data points but keep the calculated positions untouched. To prevent unnecessary calculation in such cases, we provide APIs to let users define update triggers and hint which attribute should be re-calculated given a conditional trigger.

\textbf{Per-attribute partial update:} A common use case for visualization-aided data editing is to enable interaction with one or a few data points. For example, drag and drop to edit a street segment on a map. Such interaction often results in frequent data updates as the cursor moves, even with throttling. To support efficient data handling, we added APIs to allow specifying a continuous range for partial buffer update. (So far, all discrete buffer updating use cases we encountered can be converted to use GPGPU-based editing techniques. Thus we kept a clean API to only support range-based updates.)

Familiarized with the above data wrangling strategies, users can easily integrate other tools for data processing such as layout calculation, and delegate the rendering to deck.gl for better performance and interactivity. (\textbf{E1})

\subsection{GPGPU support}
\label{sec:gpgpu}

Nowadays, even mobile devices are equipped with GPUs with computing powers more than sufficient for regular uses. On the other hand, large portion of data manipulation in data visualization is embarrassingly parallelizable. (Recall in Section \ref{sec:PIL} that when instancing a primitive, the data mapping for each datum is independent of each other.) Thus, it is natural to make use of our GPU for general purpose computations, in addition to high-performance rendering.

For example, one routine operation for geospatial visualization is to convert the input latitude and longitude of a given geolocation to screen space coordinates via Mercator projection. Such operation can be implemented in the WebGL shader to achieve a high level of parallelism for a performance boost.

Besides datum-wise operations, GPGPU is also helpful in facilitating common data processing tasks such as aggregation and filtering. The official support of Transform Feedback in WebGL2 made it even easier to create functional, off-screen layers serve as intermediate buffers to facilitie data filtering, rendering, and interactions of the final visualization.

As an example, users can create an invisible cylinder layer attached to a scatterplot layer. The instanced cylinder layer shares point positions with the scatterplot layer and renders cylinders perpendicularly facing outward from the screen. As such, the instanced cylinder layer forms a 2D Voronoi tessellation with cells encapsulating the scatterplot points. The application can then obtain the encoded color index from the cylinder when users hover over the Voronoi cells to achieve nearest point selection.

One caveat of using GPGPU in a web environment application, though, is the lack of double floating point precision support in the current WebGL shader. We have encountered cases where users need to visualize data point covering high dynamic ranges. For example, visualizing geolocations from country-level to street-level for a large number of data points. The lack of floating precision will be exaggerated at a high-zoom level, which leads to a wobbling effect and less trustworthy visual results.

To unlock this limitation, we added 64-bit floating point support in deck.gl via a collection of shader modules emulating arithmetic operations using multiple 32-bit native floating point numbers to extend the significant digits. As a result, data of high dynamic ranges (e.g. from choropleths of a continent spanning thousands of kilometers to street segments or moving objects down to centimeter level) can be processed and rendered accurately, on-the-fly.

\section{Case Studies}

In this section, we use two case studies to show how deck.gl can help scaffold visual analytics applications.

\begin{figure}[tb]
 \centering 
 \includegraphics[width=\columnwidth]{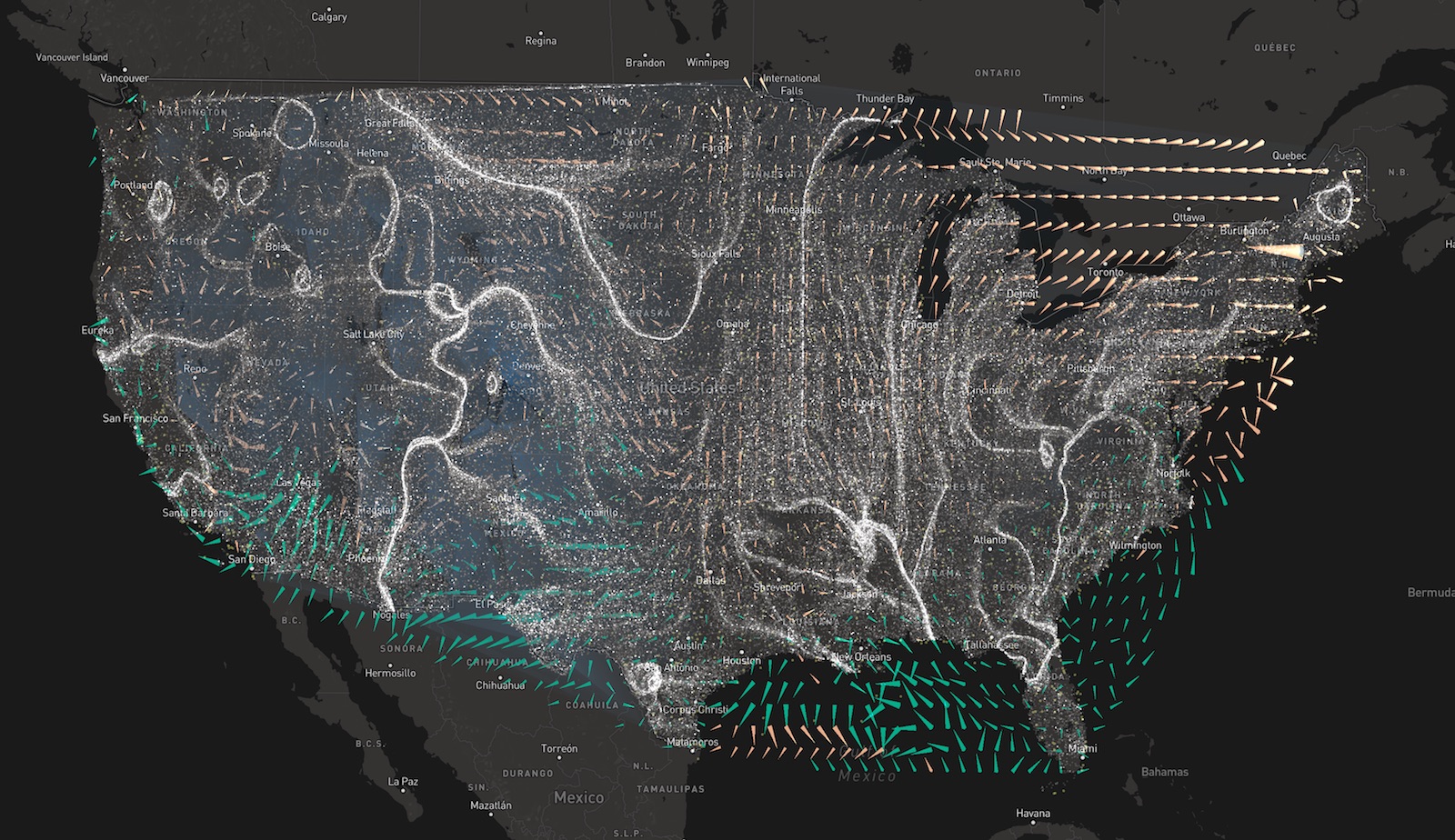}
 \caption{A screenshot of the wind simulation visualization. Three deck.gl layers are used in this demo: a functional scatterplot layer that interpolates a velocity vector from an input scattered wind station's location data; a customized vector layer that renders 3D arrows based on the interpolation result; and a point-cloud layer that also intakes the vector field as input for flow simulation.}
 \label{fig:wind}
\end{figure}

\subsection{GeoAnalytics}

GeoAnalytics is our most dominant use case as we develop deck.gl. To analyze large-scale geospatial data in the browser, we built several high-performance, data-agnostic applications powered by deck.gl. These applications are capable of rendering millions of data points and perform data transformation on-the-fly.

In the example depicted in Figure \ref{fig:wind}, we visualize a time-varying wind velocities data set from the National Oceanic and Atmospheric Administration (NOAA). The wind station locations that come with the data set are scattered across U.S. and we want to visualize what the wind flow pattern looks like at any given time stamp and location.

We use three deck.gl layers to develop this demo. Since the position of the input window stations is scattered, we first Delaunay triangulate the whole area to create a regular grid of relatively low resolution. Then, we interpolate the wind velocities in all three directions (latitude, longitude, altitude) and encode these three components as RGB colors in a functional texture layer. We then use this texture layer as input to create two on-screen layers: 1) a vector field layer that renders a customized arrow shape primitive, instanced to a regular grid that covers the whole U.S. area. For each primitive arrow, three vertex buffers (position, direction, size) are mapped from the attributes (grid point location, wind velocity, wind speed magnitude). And 2) a particle layer that also intakes the interpolated velocity field, renders to instanced point primitives randomly seeded and streamed along the vector field.

\begin{figure}[tb]
 \centering 
 \includegraphics[width=\columnwidth]{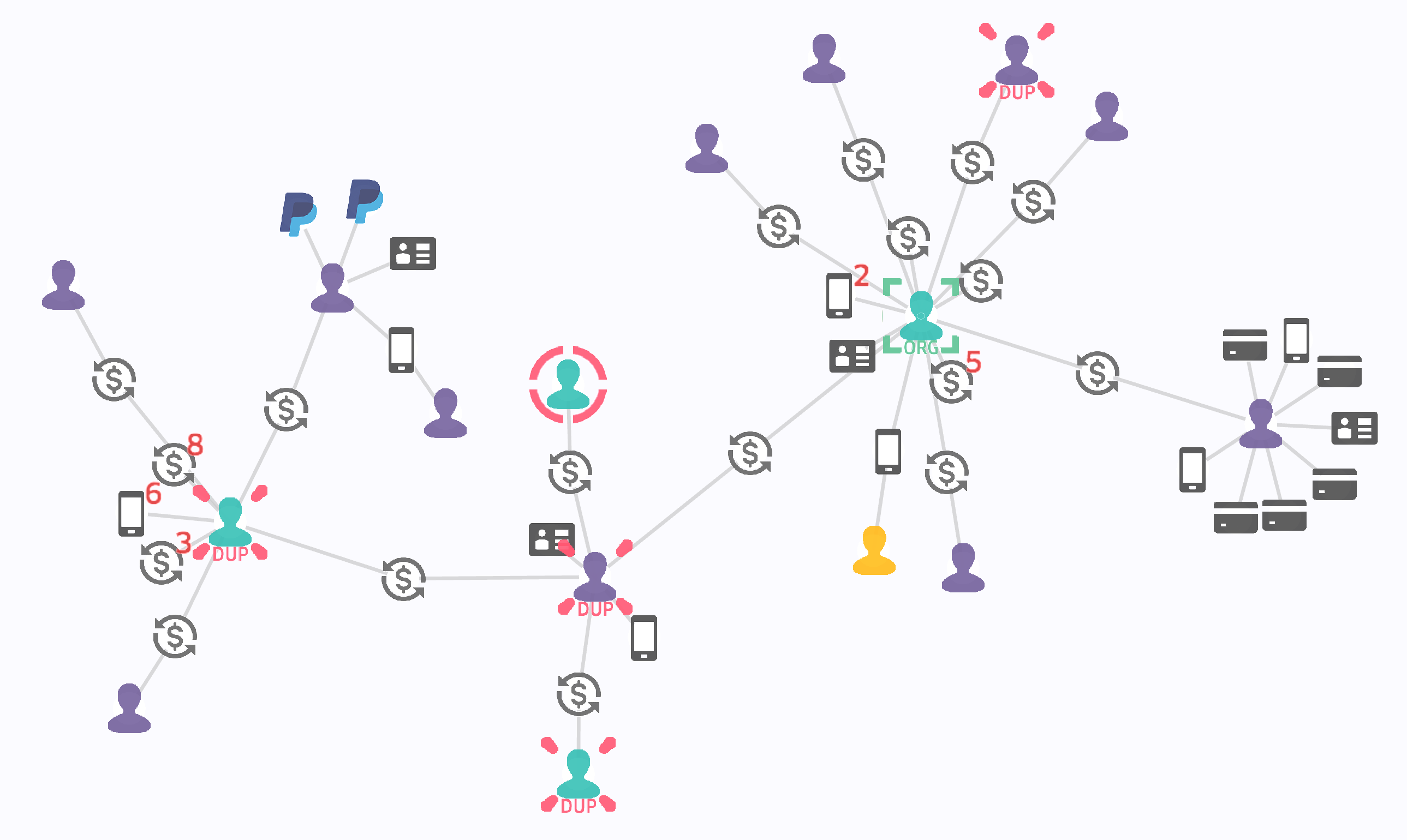}
 \caption{A network visualization created with four deck.gl layers: an icon layer renders the graph nodes; an instanced line layer renders the graph edges; one label layer renders text labels; and one customized icon layer used as decorators for semantic highlighting.}
 \label{fig:graph}
\end{figure}

\subsection{Network Analysis}

We use another example to demonstrate that the PIL paradigm is universally applicable to not only geospatial applications but also general visualization use cases. We also show how deck.gl is interoperable (\textbf{E1}) to other libraries, such as d3.js \cite{2011-d3}.

As depicted in Figure \ref{fig:graph}, in this application we model users’ trading activities within a marketplace as a network. We use a node-link diagram to reveal the interactions among entities such as user (e.g. buyer, seller), transactions, users’ payment profiles, device, etc.

To visualize such node-link diagrams at scale, we added a new graph layer by extending and compositing the existing layers in deck.gl. The graph layer is a composite layer that generates multiple sublayers. We visualize the nodes in the graph using an icon layer, which is an extension to the scatterplot layer with a rectangle primitive attached with image textures. The edges are visualized using the line layer, which can be substituted by a spline layer later when edge bundling becomes necessary. The spline layer can be considered an extension to the line layer, where each spline consists of a fixed number of line segments, instanced to approximate the curvatures.

In addition to the above two core sublayers, we also created a label layer and a decorator layer. Both of these layers are extensions of the icon layer with slightly different functionalities: while the label layer makes use of a customized texture primitive with extra styling support, the decorator layer is equipped with an internal timer for basic animation support.  We use these two layers for annotations, such as node count for collapsed super nodes and node highlight (e.g. highlight users of high similarity scores on selection) to provide extra context for data exploration.

For the graph layer, we use the d3-force utility for calculating the node positions based on its built-in n-body simulation. The d3-force utility takes two lists of objects, nodes, and edges, and injects positions based on the linkage between the nodes. Given an initial charge of force, the simulation iterates until the charge (depicted as alpha) value decreases when the simulation reaches an equilibrium state or predefined threshold. To reflect the force-directed result, we use the alpha value as the trigger to invalidate and re-generate the vertex buffer for the position attributes of the nodes and links. (Refer to Section \ref{sec:declaritive-api}, per-attribute update).

\section{Discussion}

In retrospect, we discuss a few design trade-offs we made while developing deck.gl.

\subsection{WebGL vs. SVG}

One frequently asked question we get from the visualization community is: when should one use WebGL vs. SVG for visualization? The short answer is that it depends on the number of visual elements and the level of interactivity one would like to achieve in their applications.


Take our network analysis use case for example. Based on our profiling results, the CPU time for each iteration is dominated by the update of the DOM elements rather than the calculation of the n-body simulation when the number of visual elements is small. The rendering performance degrades as the number of visual elements increases, and the sweet spot is around one thousand DOM elements. (Note: a visual element may need multiple DOM elements to encode.)

On the contrary, if we delegate the rendering to WebGL, we can render thousands or even millions of visual elements with ease. One can expect further performance boost if the heavy calculations (e.g., the Mercator projection in the GeoAnalysic cases and the n-body simulation in the force-directed graph layer) can be parallelized in the GPU. However, whether it is worth investing the time to implement complicated calculations in GPU for a large amount of data points should be evaluated on a case-by-case basis.

As a trade-off for its high performance, WebGL-based visualization has very limited native interaction support (i.e., no built-in DOM events and the need to trace target objects down in the buffer). To bridge this gap, deck.gl provides a default picking implementation and exposes the interface for both mouse and touch events via the declarative API. We design the default picking mechanism using functional layers that share the same geometric primitives with their visible duals but are colored by the object index rather than user specified colors. As such, we can easily control when to write and read from the picking buffer via a uniform switch and achieve $O(1)$ complexity for these interactions. As a result, users can expect deck.gl to work similarly to how one would handle interactions for SVG-based visualizations. 

\subsection{Framework vs. One-off Applications}

Building a framework is time-consuming and often leads to questions regarding the ROI. In the course of developing deck.gl, we are now more confident than ever before that the framework approach is beneficial in the long run. As we build new visualization solutions, we constantly review the design such that new functionalities can be distilled and merged back to push the boundary of the framework. We make use of the generalized solutions (e.g. layers in deck.gl) whenever possible, and the framework has helped us maintain a minimum overhead when scaffolding new applications.

The framework approach also helped the team scale. As new team members on board, they go over framework documents and examples and keep us aware of what is missing or confusing based on the current design. Once a new member ramps up with the framework, they are set up for success on the platform and can easily hop between different projects across teams. We argue that the same is true for users in a research institute setting, as well as for the broader visualization community.

\section{Conclusion}

In this paper, we present our motivation for developing a new data visualization framework--deck.gl--and summarize our design goals based on user feedback from the open source community. With the selected features and case studies, we demonstrate how deck.gl and the PIL paradigm can be applied to ease the creation of data-heavy visualization applications. Based on our experience, we suggest that visualization practitioners should consider WebGL-based solutions in early explorations, especially when there is little to no prior knowledge about the data. Nonetheless, we should also strive to reduce visual complexities as we gain a better understand of our data, and try to control the number of visual elements and channels used for deliverable solutions.

We hope that deck.gl, and our lessons learned while developing it, is beneficial to the visualization community.



\acknowledgments{
The author gives thank to Ravi Akkenapally, Nicolas Belmonte, Xiaoji Chen, Ib Green, Shan He, Shaojing Li, and Eric Socolofsky for their insights and all deck.gl users for providing valuable feedback to the development of deck.gl.}

\bibliographystyle{abbrv-doi-hyperref-narrow}

\bibliography{vip}

\begin{thebibliography}{1}
\renewcommand*{\sfdefault}{PTSansNarrow-TLF}

\bibitem{aframe}
https://aframe.io/.

\bibitem{react.js}
\href{https://facebook.github.io/react/}{https://facebook.github.io/react/}.

\bibitem{2011-d3}
\href{http://vis.stanford.edu/papers/d3}{M.~Bostock, V.~Ogievetsky, and
  J.~Heer}.
\newblock \href{http://vis.stanford.edu/papers/d3}{D3: Data-driven documents}.
\newblock \href{http://vis.stanford.edu/papers/d3}{{\em IEEE Trans.
  Visualization \& Comp. Graphics (Proc. InfoVis)}},
  \href{http://vis.stanford.edu/papers/d3}{2011}.

\bibitem{butler2008geojson}
H.~Butler, M.~Daly, A.~Doyle, S.~Gillies, T.~Schaub, and C.~Schmidt.
\newblock The geojson format specification.
\newblock {\em Rapport technique}, 67, 2008.

\end{thebibliography}
\end{document}